\documentclass[reprint,
nofootinbib,
longbibliography,
bibnotes,
 amsmath,amssymb, 
 prl,
]{revtex4-2}
\usepackage{amsmath,amssymb}
\usepackage{slashed,verbatim,graphicx}
\usepackage{graphicx}
\usepackage{placeins}
\graphicspath{{figures/}}
\usepackage{ytableau}
\usepackage{pstricks}
\usepackage{color}
\usepackage{tikz}
\usepackage{physics}
\usepackage[pdftex]{hyperref}
\usepackage[caption=false]{subfig}
\usepackage{float}

\newcommand{\nn}{\nonumber \\}

\newcommand{\mC}{\mathbb{C}}

\newcommand{\mc}{\mathcal}

\ytableausetup
{boxsize=0.9em}
\ytableausetup
{aligntableaux=center}

\begin{document}
\title{ Integrable Spin Chains from  large-$N$ QCD at strong coupling}
\author{David Berenstein$^{1,2,3}$ }
\author{Hiroki Kawai$^1$}
\affiliation{$^1$ Department of Physics, University of California at Santa Barbara, CA 93106\\
$^2$ 
Institute of Physics, University of Amsterdam, Science Park 904, PO Box 94485, 1090 GL Amsterdam, The Netherlands\\
$^3$ Delta Institute for Theoretical Physics, Science Park 904, PO Box 94485, 1090 GL Amsterdam, The Netherlands}
\date{\today}

\begin{abstract}
    We study a spin chain for a confining string that arises at first order in degenerate perturbation from the strong-coupling expansion of the Kogut-Susskind Hamiltonian on a square lattice in the leading large $N$ expansion. We show some subsectors are integrable and that with a relaxed constraint related to zigzag symmetry, the full spin chain is integrable in arbitrary dimensions.
\end{abstract}
\maketitle

%\section{Introduction}

The real-time dynamical evolution of quantum field theories is a very difficult theoretical  problem to tackle. Monte Carlo simulations based on path integrals fail because of the severe sign problem associated with the imaginary action.
Recent progress in quantum computing promises a solution to this simulation problem by solving a Hamiltonian evolution problem directly.
These implementations are based on the Kogut-Susskind Hamiltonian \cite{Kogut:1974ag}. A recent review can be found in \cite{Bauer:2023qgm}.
Other implementations such as quantum link models \cite{Chandrasekharan:1996ih,Brower:1997ha,Beard:1997ic,Brower:1997ha,Brower:2003vy} can be shown to be closely related to the Kogut-Susskind formulation \cite{Brower:2020huh,Berenstein:2022wlm} and differ from it by the addition of irrelevant terms.
No matter what choice is made at this stage, it is clear that with current devices, the simulation of a reasonable volume of QCD requires more resources than what are available now \cite{Klco:2021lap}.

QCD, based on the $SU(3)$ gauge group is the theory of the strong interactions \cite{Gross:1973ju,Politzer:1973fx,Gross:1973id,Gross:1974cs}.
It is a fact that representations of $SU(3)$ can have a lot of states that make the simulations hard and expensive in terms of how many qubits and quantum gates are required. 
QCD is confining and has a very rich spectrum of excitations (mesons and baryons), most of which are not calculable by lattice simulations. Progress in the direction of determining this spectrum from a first principles calculation would be very welcome. A qualitative description of these mesons is based on the idea that QCD leads to a confining string and understanding this string dynamics in detail may be approachable by an effective 1-D theory. Efforts have been made to write the effective string action of a confining flux tube such as~\cite{Polchinski:1991ax,Dubovsky:2015zey}. A 1-D theory might also be implemented on qubits more easily than a volume's worth of QCD. The question is then how to develop such a qubit model from first principles.

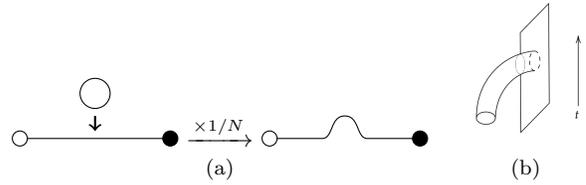
\begin{figure}[t]
    \centering
    \subfloat[]{
    \begin{tikzpicture}
        \draw(-1, 0) circle [radius = 0.1] ;
        \draw(-0.9, 0) -- (0.9, 0);
        \draw[fill] (1, 0) circle [radius = 0.1];
        \draw (0, 0.6) circle (0.2);  
        \draw[thick, <-] (0, 0.1) -- (0, 0.3);
    \end{tikzpicture}
    $\xrightarrow{\times 1/N}$
    \begin{tikzpicture}
        \draw(-1, 0) circle [radius = 0.1] ;
        \draw(-0.9, 0) --  (-0.3, 0).. controls (-0.15, 0) and (-0.2, 0.3) .. (0, 0.3) .. controls (0.2, 0.3) and (0.15, 0) .. (0.3, 0) -- (0.9, 0);
        \draw[fill] (1, 0) circle [radius = 0.1];
    \end{tikzpicture}
    \label{subfig:glueball-shoot}
    }
    \subfloat[]{
    \includegraphics[width = 0.25\linewidth]{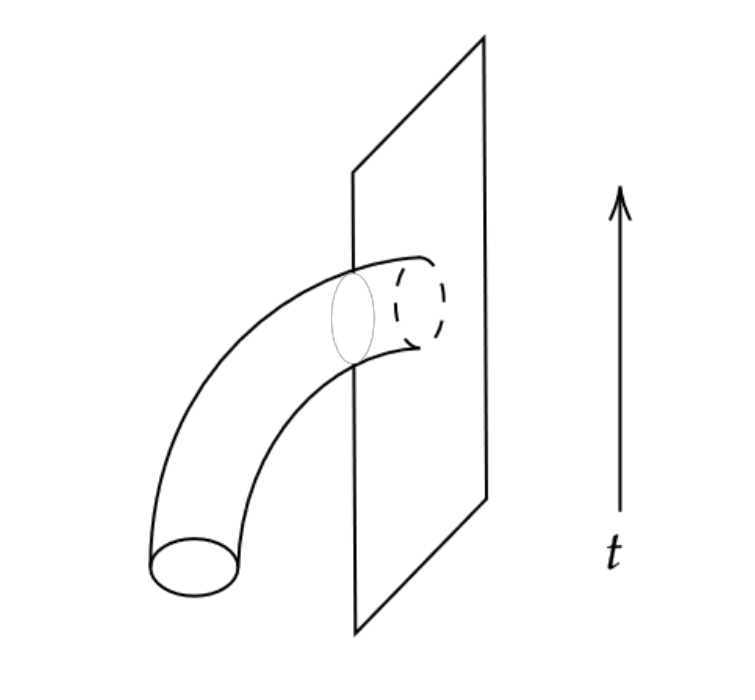}
    \label{subfig:nonplanar-worldsheet}
    }
    \caption{(a) A confining string may deform by absorbing or emitting one or more glueballs (Wilson loop excitations). This process sweeps a worldsheet geometry as (b), which is non-planar with a handle. The contributions from such processes are suppressed once we consider the large $N$ limit. }
    \label{fig:string-glueball}
\end{figure}

Our idea is to exploit the large $N$ limit of 't Hooft \cite{tHooft:1973alw}, which shows a clear connection between gauge field theories and strings. The most important example where this has been understood in detail is the AdS/CFT correspondence \cite{Maldacena:1997re}.
The dynamics of a confining string are not unitary on the worldsheet in general. 
A decay process outside of the effective $1+1$ dynamics  is possible if  a single string turns into a state with not only the string itself but also a glueball besides it. All one needs is more energy in the initial state than that of a long string plus a glueball. The virtue of the large-$N$ limit is that the amplitudes of such non-unitary processes  are non-planar and therefore suppressed (see Fig.~\ref{fig:string-glueball}).  

 We will study the large $N$ pure QCD system on a square lattice using the Kogut-Susskind Hamiltonian. 
 We will start with the strong-coupling expansion. This is  the regime where the quantum simulation is most natural.
 We will show that in perturbation theory about strong coupling, the large $N$ string leads to a family of problems  that  can be described  as spin chain Hamiltonians. 
 These are similar to the spin chains that arise in the study of the planar perturbative ${\cal N}=4 $ SYM theory \cite{Berenstein:2002jq,Minahan:2002ve,Beisert:2003tq}, which is integrable (for a review see \cite{Beisert:2010jr}). 
 The surprise in our first-order computation is that the QCD spin chain that we derive (with some additional assumptions) is actually integrable in arbitrary dimensions.
Integrability in many problems historically has been a valued source of inspiration and  led to further developments. Finding another instance of integrability is therefore important.  This integrability of the QCD string is expected to break down at higher orders in perturbation theory, but the integrable structure might be a good place to start for further computations. 
Here, we report on the integrability of the spin chain dynamics and leave other questions about the physics of the spin chains for future work.

\subsection{Plaquette actions on the strings}
The Hamiltonian formalism of the lattice gauge theory with the $SU(N)$ gauge group and coupling $g$ can be realized with the Kogut-Susskind Hamiltonian
\begin{align}
     H_{\text{KS}} = \frac{\lambda}{2N}\sum_\ell \Tr[E_\ell^2] - \frac{N}{2\lambda}\sum_P \Tr[U_P + U_P^\dagger]
\end{align}
where the first term is a sum over all the lattice links (the electric fields), and the second term is the sum over all the single plaquettes in the spatial lattice (the magnetic terms). 
The trace operator is over the color indices. The coupling constant $\lambda=g^2N$ is the t' Hooft coupling. 
In the strong-coupling limit $\lambda \gg 1$, the first term dominates and the second term can be regarded as a perturbation. 
The non-perturbed eigenbasis, i.e. the eigenbasis of the first term, consists of the ground state $\ket{\Omega}$ which has no gluon excitation.  
Excitations are generated by acting with link operators $U$ on it. Each $U$ associated with a link adds a segment starting on one vertex of the link and ending on the other that is dictated by the gauge quantum numbers. 
For example, $U^\dagger$ adds a link in the opposite direction of $U$, and the unitarity $UU^\dagger=1$ indicates that these links cancel each other.
These links must form closed directed paths for gauge invariance.
We also consider heavy quarks that allow open strings with fixed endpoints at the location of the quarks. 
Our analysis is in the bulk of the string. 
The endpoints are convenient artifacts, so as not to have to deal with the cyclic property of the trace and zero modes. It also allows certain subsectors to be easily identified.

A string is therefore a directed path from one heavy quark to an anti-quark. A path can cross itself. The large $N$ limit guarantees that these possible interconnections do not generate additional contributions, except at subleading orders in $1/N$, which we ignore.
The non-perturbed eigenbasis includes states with gluon excitations (deformations of paths) along these open strings.
The non-perturbed energies of such states are proportional to their string length, $E\sim \lambda L$. 
To compute the matrix elements of the perturbation in this basis, one needs to consider the actions of the single plaquette operators on these string states. 
There are three cases: (i) the plaquette does not share any link with the string, (ii) it shares some link(s), which have the same flux direction as the string flux, and (iii) the shared link(s) have opposite directions. 
The corrections from the first and second cases are the same as the corrections to the vacuum: they do not lead to planar corrections to the energy of the string. We focus solely on the third case. 
For the third case, the integral over a single link with opposite flux $\int d U \; {U^i}_j U^{\dagger k}\!_\ell \propto \frac{1}{N} \delta^{i}_\ell\delta^k_j$ indicates that the excitation on the shared link(s) cancels, and hence the plaquette deforms the string with the extra factor $1/N$. 
This extra factor cancels the usual $N$ factor in the Hamiltonian. The integral arises from overlap calculations.
It results in a finite contribution to the energy.
Since there are a lot of paths of the same length between a  quark and anti-quark pair, the problem of determining the effective Hamiltonian on the string is one in degenerate perturbation theory. The large $N$ limit also ensures the locality of the effective theory on the string, similar to how it works in the ${\cal N}= 4$ SYM spin chain \cite{Berenstein:2002jq}.

We label a string state by the path it takes. The string is a word with instructions on what is the next step: up, down, left, right, etc. Each of these is a letter of the alphabet. 
An action of the plaquette changes the shape of the path. We only consider those changes that contribute to the degenerate perturbation theory. Therefore, the change of shape keeps the number of 
letters fixed. The $UU^\dagger=1$ constraint is implemented by stating that a path can not double back on itself immediately. That is, combinations such as  $\dots ud\dots$, or $\dots l r\dots$ are forbidden. This is also called zigzag symmetry and is related to the reparametrization invariance on the worldsheet~\cite{Polyakov:1997tj}.

\subsection{Diagonal string}
The first case we consider is the strings with endpoints diagonal from each other. 
One possible excitation is along the path as the black thick path in the left figure of Fig.~\ref{fig:diagonal-plaquette} with length $L$. 
It contains an equal number ($L/2$) of horizontal and vertical links and only has excitations of two letters words ($ur$ for example). 
Let us call this path a diagonal string, $\Gamma$. 
The single plaquette operator sharing exactly two links with $\Gamma$ as the red plaquette in Fig.~\ref{fig:diagonal-plaquette} generates another string excitation generated with the same string length and the same positions of the quark-antiquark endpoints.
Hence, the excitation along $\Gamma$ is not a unique lowest energy excitation but rather is degenerate with the states generated by a series of this kind of plaquette actions. 
The Hilbert space, which we call $\mc H_{\Gamma}$, spanned with these degenerate states has a dimension of ${ L \choose L/2}$.

\begin{figure}[ht]
    \centering
    \begin{figure}[H]
    \centering
    \begin{tikzpicture}[baseline = (current bounding box.center)]
        \draw[step=0.5,gray,very thin] (-1.2,-1.2) grid (1.2,1.2);
        \draw[very thick] (-1, -1) -- (-0.5, -1)--(-0.5, -0.5)--(0, -0.5)--(0, 0) --(0.5, 0) -- (0.5, 0.5)--(1, 0.5)--(1, 1); 
        \draw[very thick, red] (-0.5, -0.5) -- (-0.5, 0) -- (0, 0);
        \draw[very thick, red, dashed] (-0.5, -0.5) -- (0, -0.5) -- (0, 0);
    \end{tikzpicture}
    $\large \rightarrow$
    \begin{tikzpicture}[baseline = (current bounding box.center)]
        \draw[step=0.5,gray,very thin] (-1.2,-1.2) grid (1.2,1.2);
        \draw[very thick] (-1, -1) -- (-0.5, -1)--(-0.5, -0.5)--(-0.5, 0)--(0, 0) --(0.5, 0) -- (0.5, 0.5)--(1, 0.5)--(1, 1); 
    \end{tikzpicture}
\end{figure}
    \caption{The diagonal string $\Gamma$ with $L/2$ horizontal links and $L/2$ vertical links (thick black path).
    If one acts the single plaquette operator (red) sharing two links (red dotted lines) on the diagonal string state, it changes the state to another with the same string length and the endpoints. }
    \label{fig:diagonal-plaquette}
\end{figure}

The first-order perturbation to the strong-coupling limit $\lambda \gg 1$ can be found by diagonalizing the ${ L \choose L/2}\times { L \choose L/2}$ matrix 
\begin{align}
    W_{ij} = \mel{\Gamma_i}{\left(-\frac{N}{2\lambda} \sum_P \Tr[U_P + U_P^\dagger] \right)}{\Gamma_j}
\end{align}
where $\ket{\Gamma_i} \in \mc H_\Gamma$. 
The action of a single plaquette operator sharing two of its edges with the string can be seen as flipping neighborhood vertical and horizontal links. 
Regarding a vertical link to be an up-spin $\uparrow$ and a horizontal link to be a down-spin $\downarrow$, the plaquette operator flips the spins of two neighborhood spin-$1/2$ sites with opposite spins, i.e. 
\begin{align}
    \Tr[U_P+U_P^\dagger] \sim \frac{1}{N}(\sigma^+_x \sigma^-_y + \sigma^-_x \sigma^+_y)
\end{align}
The whole string is the spin-$1/2$ chain with length $L$. 
Hence, the action of the magnetic perturbation $-\frac{N}{2\lambda} \sum_P \Tr[U_P + U_P^\dagger]$ is the same as the XX model Hamiltonian  
\begin{align}\label{eq:H-XX}
     H = -\frac{1}{2\lambda} \sum_{j = 1}^{L} (\sigma^+_ j\sigma^-_{j+1} + \sigma^-_j \sigma^+_{j+1})
\end{align}
This Hamiltonian conserves the magnetization, and one can decompose the Hilbert space to the sectors with different magnetization values as 
$
     H_{\text{XX}} = \bigoplus_{m = 0}^{L} \mc H_m
$
where $\mc H_m$ is the sector having $m$ down spins, whose dimension is ${ L \choose m} $. 
The different sectors correspond to the string sectors with different endpoints with the string length fixed to be $L$. 
The sub-Hilbert space we are interested in, $\mc H_{\Gamma}$, is exactly the $\mc H_{L/2}$ sector. 
The first-order perturbation matrix $W_{ij}$ is exactly the block of the XX model Hamiltonian in $\mc H_{L/2}$. 
The XX model is trivially integrable given that it is equivalent to two copies of a free fermion theory. 
Kogut et al. found this free fermion description in~\cite{Kogut1981}.
Nearest-neighbor spin chains such as these are easily implementable in quantum computers. 
The fact that this subsector is integrable should be no surprise: spin chains only with nearest-neighbor interactions preserving the spin $\sigma_z$ that are parity conserving (treat $\uparrow\downarrow$ equivalently) are all members of the $XXZ$ family. 

\subsection{Straight string}

\begin{figure}[t]
    \centering
    \subfloat[]{\label{fig:straight}\includegraphics[width = 0.45\linewidth]{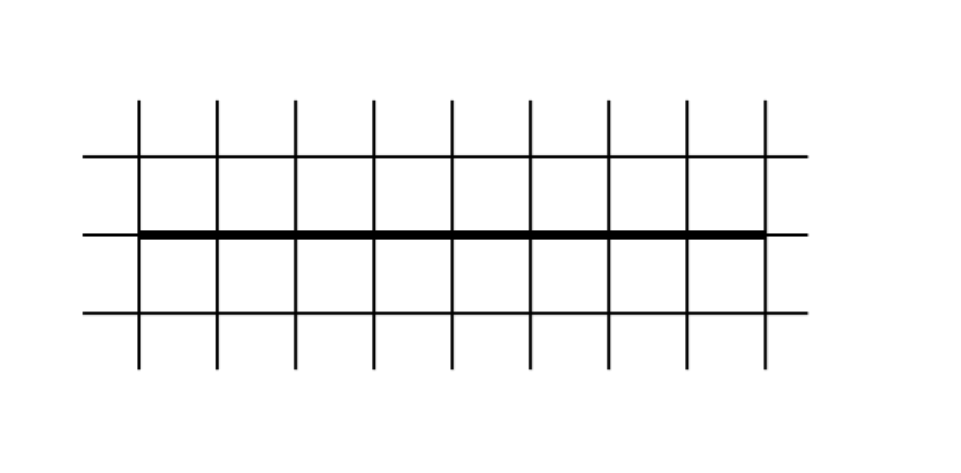}}
    \subfloat[]{\label{fig:straight-correction}\includegraphics[width = 0.45\linewidth]{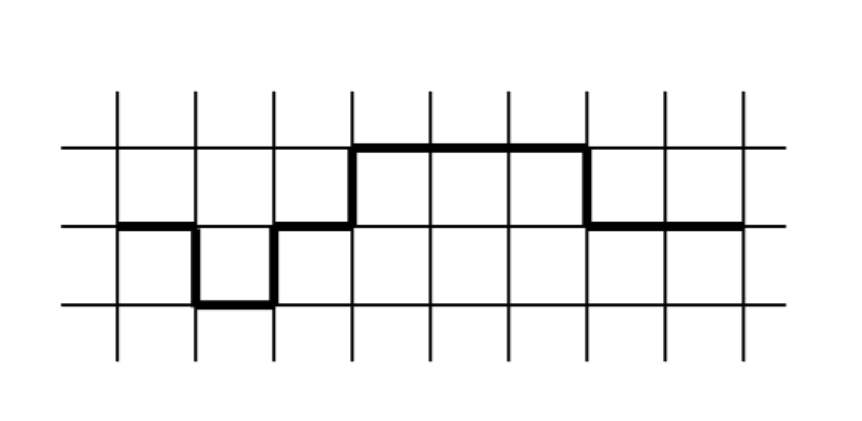}}
    \caption{(a) The straight string $\Tilde \Gamma$, the ground state. 
    (b) Paths $\Tilde \Gamma$ coming from the configurations with some misalignments treated as excitations (the vertical link excitations). }
\end{figure}

We now turn to the second case of the string excitation along the path with the endpoints horizontally aligned along one of the spatial lattice edge directions. 
We call the shortest path connecting them $\Tilde \Gamma$ (Fig.~\ref{fig:straight}) and consider the other shapes with the same endpoints as excitations on top of the ground state. 
To characterize string configurations, we express them with three letters $udr$, being careful to disallow $ud$ and $du$ words. When we act with the plaquette, one easily sees that $ur$ and $dr$ pieces of the Hamiltonian look the same as those of the diagonal string, which allows swapping the letters. 
The new wrinkle is that swapping the order $urd$, for example, can land one in the forbidden configuration $ud$.
In that sense, those possibilities must be projected out. Such projections effectively turn the problem into one where the swap of letters is allowed depending on extra letters, a three- or four-letter problem. The upshot is that at this order in perturbation theory, $ud$ letters can not swap position. 
This property is similar to that of the eclectic and hyper-eclectic spin chain where some letters (states) act sometimes as walls \cite{Ipsen:2018fmu,Ahn:2020zly}.

It is convenient to pass to a new basis, where we treat $ud$ as walls, and use an occupation number basis for $r$ in between each pair of walls. This is identical to the  treatment employed to describe the XXX spin-chain with variable length in~\cite{Berenstein:2005fa}. 
In this basis, the Hamiltonian for the perturbation becomes
\begin{equation}
   H=- \frac{1}{2\lambda} \sum (c_i^\dagger c_{i+1} +c_i c^\dagger_{i+1})
\end{equation}
where each of the $c_i$ are the Cuntz oscillators satisfying $c_i c_i^\dagger=1$. Their action is $c^\dagger \ket n= \ket{n+1}$ and $c\ket 0=0$ for each Fock space. We restrict the Hilbert space by requiring that the occupation number between any walls that change direction $ud$ or $du$ is greater than or equal to one. It turns out that this restriction leaves the Cuntz action unchanged on the relevant states as if the occupation one state had occupation 
number zero: the algebra does not change if we do a shift by one plus a projection.  In that sense, we can relabel the states with a modified Cuntz oscillator at these sites that ignores that constraint (it becomes an implicit padding of the words). That is, we substitute $\ket 1 \rightarrow \ket 0: dru \rightarrow du, \; \ket 2 \rightarrow \ket 1: drru\rightarrow dru$, etc. Basically, we are now allowing the $ud$ to be neighbors without any restrictions. The new labels 
now allow us to describe the whole problem as a nearest neighbor spin chain without extra projections. 

The result is a simple Hamiltonian where only swaps $d \leftrightarrow r$ and $u\leftrightarrow r$ and their reverse swaps are allowed. 
All other terms in the Hamiltonian vanish. We can map these to a spin one system as $r\to\ket 0$, $u\to \ket +$, $d\to \ket -$. 
The Hamiltonian is equivalent to the following spin-chain Hamiltonian acting on this Hilbert space $\mc H$:  
\begin{align}
     H = -\frac{1}{2\lambda}  \sum_{j = 1}^L \left(e^1_j f^1_{j+1} + f^1_j e^1_{j+1} + e^2_j f^2_{j+1} + f^2_j e^2_{j+1} \right)
\end{align}
with the generators of $\mathfrak{sl}(3, \mC)$:
\begin{align}
    h^1 &= 
    \begin{pmatrix}
        1 & 0 & 0\\ 
        0 & -1 & 0\\
        0 & 0 & 0
    \end{pmatrix}, 
    \; 
    h^3 = 
    \begin{pmatrix}
        0 & 0 & 0\\ 
        0 & 1 & 0\\
        0 & 0 & -1
    \end{pmatrix}
    \nn
    e^1 &= 
    \begin{pmatrix}
        0 & 1 & 0\\ 
        0 & 0 & 0\\
        0 & 0 & 0
    \end{pmatrix}, 
    \; 
    e^2 = 
    \begin{pmatrix}
        0 & 0 & 0\\ 
        0 & 0 & 0\\
        1 & 0 & 0
    \end{pmatrix}
    , \; 
    e^3 = 
    \begin{pmatrix}
        0 & 0 & 0\\ 
        0 & 0 & 1\\
        0 & 0 & 0
    \end{pmatrix}
    \nn
    f^1 &= 
    \begin{pmatrix}
        0 & 0 & 0\\ 
        1 & 0 & 0\\
        0 & 0 & 0
    \end{pmatrix}, 
    \; 
    f^2 = 
    \begin{pmatrix}
        0 & 0 & 1\\ 
        0 & 0 & 0\\
        0 & 0 & 0
    \end{pmatrix}, 
    \; 
    f^3 = 
    \begin{pmatrix}
        0 & 0 & 0\\ 
        0 & 0 & 0\\
        0 & 1 & 0
    \end{pmatrix}
\end{align}
The subscript $j$ specifies that the operator is acting on the $j$-th spin site. 
Each term exchanges the neighbor $\ket 0$ and either $\ket{+}$ or $\ket{-}$. 
$\ket{+}$ and $\ket{-}$ cannot pass through each other.
This model has an $SU(2) \times U(1)_i$ symmetry generated by $ h = \sum_{j=1}^L h^3_j$, $ e = \sum_{j=1}^L e^3_j$, and $ f = \sum_{j=1}^L f^3_j$ for the $SU(2)$, and $ i = \sum_{j=1}^L i^3_j$ for $U(1)_i$ with 
\begin{align}
    i^3 = \begin{pmatrix}
        0 & 0 & 0\\
        0 & 1 & 0\\
        0 & 0 & 1
    \end{pmatrix}
\end{align}
One can interpret the $SU(2)$ part as the invariance of the system under exchanges of $\ket{+}$ $\leftrightarrow $ $\ket{-}$. 
The extra symmetry suggests that the model might be integrable.
Indeed, the Hamiltonian can be constructed (up to the coupling $-1/2\lambda$ factor) from the R-matrix 
\begin{widetext}
\begin{align}
   R(\mu) 
    = 
\begin{pmatrix}
i \cosh (\mu) & 0 & 0 & 0 & 0 & 0 & 0 & 0 & 0 \\
 0 & \sinh (\mu) & 0 & i & 0 & 0 & 0 & 0 & 0 \\
 0 & 0 & \sinh (\mu) & 0 & 0 & 0 & i & 0 & 0 \\
 0 & i & 0 & \sinh (\mu) & 0 & 0 & 0 & 0 & 0 \\
 0 & 0 & 0 & 0 & i \cosh (\mu) & 0 & 0 & 0 & 0 \\
 0 & 0 & 0 & 0 & 0 & 0 & 0 & i \cosh (\mu) & 0 \\
 0 & 0 & i & 0 & 0 & 0 & \sinh (\mu) & 0 & 0 \\
 0 & 0 & 0 & 0 & 0 & i \cosh (\mu) & 0 & 0 & 0 \\
 0 & 0 & 0 & 0 & 0 & 0 & 0 & 0 & i \cosh (\mu) \\
\end{pmatrix}
\end{align}
\end{widetext}
which satisfies the Yang-Baxter equation. This is done by expanding the transfer matrix associated to $R$ around $\mu=0$ as the spectral parameter (see for example \cite{Retore:2021wwh}).  The full problem is therefore solvable by Bethe ansatz. This model has appeared before in \cite{doi:10.1142/S0217751X92003458,Alcaraz:1992zc,Maassarani:1997kon} where it was also shown to be integrable. Our important result is that the three state spin chain with constraints can be converted to a nearest neighbor spin chain without constraints, which happens to be a known integrable model.

This is suggestive that should consider the equivalent spin chain for Yang-Mills in higher dimensions ignoring altogether (one could say relaxing) the zigzag constraints. One allows swaps of all pairs of letters, except the ones that are paired (ud, lr, etc).
It is straightforward to check that all of these Hamiltonians with the relaxed zigzag constraint are integrable, with an R-matrix that is similar to the one above. Swaps have the $\sinh(\mu)$ terms with an off-diagonal $i$, and non-swaps have the off-diagonal $i \cosh(\mu)$ blocks.
Identical letters result in a $1\times 1$ diagonal block  $i \cosh(u)$.
Whether the full spin chain {\em with all restrictions included} is also integrable we cannot yet ascertain. One can also think that the zigzag constraint is a UV problem at the lattice scale and that the IR of the effective string doesn't care about it. One can also go via another route: ignore the constraint and add an energy penalty for forbidden configurations to get a nearest-neighbor spin chain. 

We leave a more detailed study of the physics of this system in our forthcoming paper.
Notice that integrability in this case is about the bulk Hamiltonian (the closed strings). We have said nothing about if and how the boundary conditions are compatible with this integrability. In our case, they were simple ways to get restrictions to subsectors with fewer letters.
A similar reduction in letters would be obtained by considering the Eguchi-Kawai reduction \cite{Eguchi:1982nm} instead.

It is well known that the strong-coupling expansion of QCD is not relativistic and is trivially confining (very similar in spirit to Wilson's argument \cite{Wilson:1974sk}). One needs to approach the roughening transition point to get to the correct theory of the confining string. Understanding the approach to this transition on the string worldsheet starting from a Hamiltonian formulation at strong coupling is interesting and worth exploring. Our results show that this is indeed the case. 

\begin{acknowledgments}
We would like to thank Richard Brower, Sergei Dubovsky, Ana Retore, and Matthias Staudacher for many discussions.  We would also like to thank Hosho Katsura for correspondence.  This work resulted from the KITP workshops on ``Confinement, Flux Tubes, and Large N" and ``Integrability in String, Field, and Condensed Matter Theory". Research supported in part by the Department of Energy under Award No. DE- SC0019139. This research was also supported in part by the National Science Foundation under Grants No. NSF PHY-1748958 and PHY-2309135. DB is also  supported in part by the Delta ITP consortium, a program of the Netherlands Organisation for Scientific Research (NWO) funded by the Dutch Ministry of Education, Culture and Science (OCW)
\end{acknowledgments}

\bibliographystyle{apsrev4-2}
\bibliography{ref}

\end{document}